\documentstyle[12pt,epsfig]{article}

\newcommand{\be}{\begin{equation}}
\newcommand{\ee}{\end{equation}}
\def\bea{\begin{eqnarray}}
\def\eea{\end{eqnarray}}
\def\nn{\nonumber \\}

\input psfig

\begin{document}

\date{}
\title{
{\large\rm DESY 97-066}\hfill{\large\tt ISSN 0418-9833}\\
{\large\rm ZU-TH 9/97}\hfill\vspace*{0cm}\\
{\large\rm April 1997}\hfill\vspace*{2.5cm}\\
Constraints on the Universal Contact Interaction}
\author{W. Buchm\"uller\\
{\normalsize\it Deutsches Elektronen-Synchrotron DESY, 22603 Hamburg, Germany}
\\[.2cm]
and\\[.2cm]
D. Wyler\\
{\normalsize\it Institut f\"ur Theoretische Physik, Universit\"at Z\"urich,
Z\"urich, Switzerland }
\vspace*{2cm}\\                     
}                                                                          

\maketitle  
\begin{abstract}
\noindent
Forces beyond those of the standard model may manifest themselves at low 
energies as four-fermion contact interactions. If these new forces are
independent of colour and flavour quantum numbers including baryon and
lepton number, then  all low energy constraints,
arising from quark-lepton universality, flavour-changing neutral currents
and atomic parity violation are evaded. This is due to the global U(45)
symmetry which the standard model exhibits in the limit of vanishing gauge 
and Yukawa couplings. The corresponding contact interaction is a unique  
current-current interaction. Constraints from LEP2 imply that this universal
contact interaction cannot be the origin of the recently observed high-$Q^2$ 
events at HERA.
\end{abstract} 
\thispagestyle{empty}

\newpage                                             
The success of the standard model of strong and electroweak interactions 
is partially related to its approximate global symmetries which `explain' 
certain numerical relations. One example is the `custodial' SU(2)$_R$ symmetry 
\cite{sikivie} under which right-handed fermions and the Higgs field 
$\Phi=(\varphi,\tilde{\varphi})$, with 
$\tilde{\varphi}^i=\epsilon^{ij}\varphi^*_j$, transform as doublets.
In the case of unbroken SU(2)$_R$ the masses of up- and down-type quarks
are equal and the $\rho$-parameter equals one. If, furthermore, the masses 
of all up- and down-type quarks are equal, one has an additional U(n$_f$) 
family symmetry (n$_f$=3) under which left- and right-handed 
quarks transform into each other. This implies a perfect GIM cancellation
\cite{glashow}, i.e., the absense of flavour changing couplings of the
$Z$-boson to any order in perturbation theory.

The observed suppression of flavour changing neutral currents imposes strong 
constraints on extensions of the standard model (see, e.g., \cite{buwy}). An 
interesting example is the rare kaon decay mode K$_L\rightarrow e^-\mu^+$.
In extensions of the standard model with couplings of quark-lepton pairs to 
new states electron and muon number are in general not separately conserved.
In this case the (unobserved) $K_L$ decay process can be forbidden by the 
global symmetry U(4n$_f$)$\supset$SU(4)$\otimes$U(n$_f$), where 
SU(4) is the Pati-Salam group containing SU(3) colour and the U(1) of 
lepton number, and U(n$_f$) is a family symmetry \cite{greenberg}. 
Alternatively, the dangerous kaon decay mode may be prevented by means of
a separate family symmetry for quarks and leptons, i.e., by the
symmetry group U$_q$(n$_f$)$\otimes$U$_l$(n$_f$)\cite{suzuki}. 

Imposing SU(2)$_L\otimes$SU(2)$_R\otimes$U(4n$_f$) as an approximate global 
symmetry, broken only by gauge interactions, requires a right-handed neutrino,
and it reduces all Yukawa couplings of the standard model to a single one. 
Setting this Yukawa coupling as well as the gauge couplings equal to
zero yields the largest possible global symmetry, U(48), or, in the absense
of right-handed neutrinos, U(45). In this limit all fermions are treated 
completely equal.

Is it conceivable that the excess of high-$Q^2$ events in $e^+p$ deep-inelastic
scattering, which has been reported by the collaborations H1 \cite{h1} and 
ZEUS \cite{zeus} at HERA, is due to an electron-quark contact interaction 
\cite{ruckl,schrempp}. Recent analyses show that such an interpretation 
is consistent with results on Drell-Yan lepton-pair production at the Tevatron 
as well as bounds on the cross section $e^+e^-\rightarrow hadrons$ obtained at 
LEP2 \cite{altarelli,zerwas,babu,barger}. A severe constraint on allowable
contact interactions is imposed by the low energy precision measurements 
of atomic parity violation \cite{langacker}. As recently pointed out,
also this constraint can be met by a symmetry, the SU(12) invariance with 
respect to transformations among the different quark states $u_L, d_L, u_R^c, 
d_R^c$ of the first generation \cite{nelson}.

The constraints from rare processes represent a severe problem for any
interpretation of the HERA data in terms of physics beyond the standard model.
In order to avoid them, one is naturally led to consider
forces which are independent of the colour and flavour of the known
quarks and leptons. An example is a strong new U(1) gauge interaction
which may lead to a plethory of new resonances, including leptoquark
states \cite{babu}. In this note we consider constraints on
contact interactions which are universal for all fermions and therefore
invariant under the largest possible approximate global symmetry of the 
standard model, i.e., the group U(45). 

The 45 Weyl fermions of the standard model are, in obvious notation, 
represented by the spinor,
\be
\Psi_L^A = \left(q_L, u_R^c, d_R^c, l_L, e_R^c \right)\, ,
\ee
where we have dropped colour, weak isopin and generation indices. U(45)
invariant four-fermion contact interactions can be constructed from the
scalar and vector densities 
\be
\bar{\Psi}_L^{cA}\Psi_L^B\, ,\quad \bar{\Psi}_{LA} \gamma_{\mu}\Psi_L^B\, .
\ee
One easily verifies that all invariants which one may naively write
down are, via a Fierz transformation and group theoretical 
identities, related to the following unique
contact interaction,
\be\label{contact}
{\cal L}_U = \epsilon {2\pi\over \Lambda^2}\ J^{\mu}J_{\mu}\, ,
\ee
where $\epsilon=\pm 1$ and
\be\label{current}
J_{\mu} = \bar{u}\gamma_{\mu}\gamma_5 u + \bar{d}\gamma_{\mu}\gamma_5 d + 
          \bar{e}\gamma_{\mu}\gamma_5 e + \bar{\nu}_L\gamma_{\mu}\nu_L\, .
\ee
Here we have used the identity $\bar{\chi^c}\gamma_{\mu}\chi^c = -
\bar{\chi}\gamma_{\mu}\chi$, and the sum over all generations is understood.
Note, that the current $J_{\mu}$ is purely axial except for the neutrino 
contribution where the right-handed partner is absent in the standard model. 
This special form can be easily understood. The maximal symmetry U(45)
implies that the contact interaction between all fermion states of the
same chirality is identical. If two fermion states can be identified as the 
left-handed and right-handed components of a massive Dirac fermion, only the 
linear combination $\bar{f}_L\gamma_{\mu}f_L + \bar{f}^c_R\gamma_{\mu}f^c_R$
can couple to other fermion states. This linear combination corresponds to
an axial current. For a vector current the relative sign would be opposite. 

The universal contact interaction ${\cal L}_U$ is clearly compatible with 
the SU(5) unification of gauge interactions hinted at by the standard model 
and its supersymmetric extension. Demanding further the SU(12) symmetry which
guarantees that the constraint from atomic parity violation is fulfilled
\cite{nelson}, one is lead to the maximal symmetry U(15) for the first family.
Suppression of the decay $K_L \rightarrow e^-\mu^+$ appears to require
at least the family symmetry U(n$_f$), n$_f$=3. One easily verifies that the 
most general contact interaction which is invariant under U(15)$\otimes$U(3) 
is also invariant under the larger symmetry U(45). Hence, it appears difficult
to avoid this large global symmetry for a contact interaction which is
compatible with the unification of gauge interactions and which satisfies all
low energy constraints without fine tuning of parameters. 

It is remarkable that the contact interaction ${\cal L}_U$ is essentially
invisible in rare processes. It conserves parity for the interactions of
charged particles, hence it is unconstrained from atomic parity violation 
measurements. Further, $J_{\mu}$ is a universal and flavour diagonal 
neutral current; consequently contraints from quark-lepton universality
in charged current interactions or bounds on flavour changing neutral
currents do not apply. There is no contribution to the leptonic decay
of $\pi^0$ as $J_{\mu}$ is an isoscalar current. Also isoscalar meson
decays yield no relevant constraint: the contribution of the contact
interaction to $\eta\rightarrow \mu^+\mu^-$ is suppressed since $\eta$
is predominantly an SU(3) octet; 
$\eta'$, on the other hand, is very broad and the decay $\eta'\rightarrow
e^+e^-$ is chirally protected, i.e., $\Gamma \propto m_e^2$.

To see explicitly, how weak the low energy constraints on the contact
interaction ${\cal L}_U$ are, consider the decay 
$\eta \rightarrow \mu^+\mu^-$ which gives the strongest bound.
Since the experimental, and even more so the theoretical error on the decay 
rate \cite{herczeg} is of the same order as the decay rate itself,
one can obtain a rough estimate of the bound on $\Lambda$ by demanding
that the contact interaction ${\cal L}_U$ does not yield a decay rate
larger than the measured one. The decay rate is given by
\be
\Gamma_U(\eta\rightarrow \mu^+\mu^-) \simeq {3\over 2\pi} \sin^2{\Theta_P}
\left({4\pi\over \Lambda^2}\right)^2 f_{\eta'}^2 m_{\mu}^2 m_{\eta}
\left(1-{4m_{\mu}^2\over m_{\eta}^2}\right)^{1/2}\ .
\ee
Here $f_{\eta'}$, $m_{\mu}$, $m_{\eta}$ and ${\Theta_P}$ are the 
$\eta'$-decay constant, muon mass, $\eta$-mass and $\eta$-$\eta'$ mixing angle, 
respectively. Demanding $\Gamma_U < \Gamma_{exp}$ and using the measured
values for the various parameters \cite{rpp}, one obtains the lower 
bound on the mass scale $\Lambda$,
\be
\Lambda > 15\ \mbox{GeV}\, .
\ee 
This bound is two orders of magnitudes smaller than typical values obtained
from high-energy experiments! 

The universal contact interaction ${\cal L}_U$ does contribute to
deep-inelastic scattering via the contact term
\be
{\cal L}_{eq} = \epsilon{4\pi\over \Lambda^2}\ \bar{e}\gamma^{\mu}\gamma_5 e\
                \left(\bar{u}\gamma_{\mu}\gamma_5 u +
                      \bar{d}\gamma_{\mu}\gamma_5 d \right)\, .
\ee
In the usual notation,
\be
{\cal L}_{CI} = \sum_{i,j=L,R;\ q=u,d}\eta^q_{ij}\ \bar{e}_i\gamma^{\mu}e_i\
                                   \bar{q}_j\gamma_{\mu}q_j\, ,
\ee
the contact interaction ${\cal L}_{eq}$ corresponds to the choice
\bea
\eta^u_{LL}&=&\eta^u_{RR}=-\eta^u_{RL}=-\eta^u_{LR}=
\epsilon{4\pi\over \Lambda^2}\, ,\\ \nn
\eta^d_{ij}&=&\eta^u_{ij}\, .
\eea
At large values of $Q^2$, where an excess of events has been observed at
HERA, the contribution of the LL- and RR-terms is small compared to those
of the RL- and LR-terms. Also the contribution from down-quarks is small
compared to the one from up-quarks. From the analysis in \cite{barger} 
it is then clear
that a good description of the HERA data is obtained for the choice
$\epsilon=-1$ and $\Lambda\simeq 3$ TeV. 

The universal contact interaction ${\cal L}_U$ also contributes to
purely leptonic processes,
\be\label{lepcon}
{\cal L}_{el} = \epsilon{4\pi\over \Lambda^2}\ \bar{e}\gamma^{\mu}\gamma_5e\
\left({1\over 2}\bar{e}\gamma_{\mu}\gamma_5e 
      + \bar{\mu}\gamma_{\mu}\gamma_5\mu + \bar{\tau}\gamma_{\mu}\gamma_5\tau
 \right)\, .
\ee        
This contact interaction, however, is strongly contrained by data from
LEP2. An analysis of the OPAL collaboration yields for a universal leptonic 
axial-current current interaction ($\epsilon=-1$, i.e., $AA^-$ \cite{opal}),
\be
\Lambda > 5\ \mbox{TeV}\, ,
\ee
which is significantly larger than the value needed to explain the HERA data.

We conclude that the universal contact interaction ${\cal L}_U$ cannot
be the origin of the observed excess of high-$Q^2$ events at HERA. 
If due to new physics, sizeable quark-lepton symmetry breakings or new 
particles must exist.

We wish to thank P.~Herczeg, F.~Schrempp and P.~M.~Zerwas for helpful 
discussions.


\end{document}